2020

# Lotka's Law and Authorship Distribution in Coronary Artery Disease Research in South Africa

Muneer Ahmad
*Annamalai University*, muneerbangroo@gmail.com

Dr. M.Sadik Batcha
*Annamalai University*, msbau@rediffmail.com

# Lotka's Law and Authorship Distribution in Coronary Artery Disease Research in South Africa

Muneer Ahmad[1] Dr. M Sadik Batcha[2]

[1]*Research Scholar, Department of Library and Information Science, Annamalai University, Annamalai nagar, muneerbangroo@gmail.com*
[2]*Research Supervisor & Mentor, Professor and University Librarian, Annamalai University, Annamalai nagar, msbau@rediffmail.com*

**Abstract**

Coronary artery disease (CAD) is a major cause of death and disability in developed countries. Although CAD mortality rates worldwide have declined over the past 4 decades, CAD remains responsible for approximately one-third or more of all deaths in individuals over age 35, and it has been estimated that nearly half of all middle-aged men and one-third of middle aged women in the United States will develop clinical CAD. The present paper attempts to check the applicability of Lotka's Law on South African publication on Coronary artery disease research. The study lights on Lotka's empirical law of scientific productivity, i.e., Inverse Square Law, to measure the scientific productivity of authors, to test Lotka's Exponent value and the K.S test for the fitness of Lotka's Law.

## 1. Introduction

*Coronary artery disease (CAD)*-Often called coronary heart disease or CHD, is in general used to refer to the pathologic practice affecting the coronary arteries (typically atherosclerosis). CAD includes the diagnoses of angina pectoris, myocardial infarction (MI), silent myocardial ischemia, and CAD mortality that result from CAD. Hard CAD endpoints normally include MI and CAD death. The term CHD is frequently used interchangeably with CAD. *CAD death-* includes unexpected cardiac death (SCD) for conditions when the death has occurred within 24 hours of the immediate onset of symptoms, and the term non - SCD applies when the time course from the clinical appearance until the time of death exceeds 24 hours or has not been explicitly recognized (Lemos & Omland, 2018).

Coronary Artery disease is one of the foremost causes of death in industrialized countries and is accountable for one out of every six deaths in the United States. Amazingly, coronary artery disease is also mostly preventable. The pathogenesis is intricate and involves many different pathways that are only incompletely unstated. In recent times, the morbidity and mortality of ischaemic heart disease have considerably enhanced because of development in therapeutic strategies, represented by the spectacular growth of percutaneous coronary intervention (Squeri, 2012).

## 2. Review of Literature

The frequency distribution of scientific productivity (Lotka, 1926), pertaining to scattering of literature (Bradford, 1950), and concerning word frequency in the text (Zipf, 1949) are the three basic Laws of Bibliometrics. Lotka in his paper on "Analysis of the number of publications in Chemical Abstracts from 1907 to 1916 on frequency distribution of scientific productivity" proposed an inverse square law of scientific productivity.

(Ahmad, Batcha, & Jahina, 2019) quantitatively measured the research productivity in the area of artificial intelligence at global level over the study period of ten years (2008-2017). The study acknowledged the trends and features of growth and collaboration pattern of artificial intelligence research output. Average growth rate of artificial intelligence per year increases at the rate of 0.862. The multi-authorship pattern in the study is found high and the average number of authors per paper is 3.31. Collaborative Index is noted to be the highest range in the year 2014 with 3.50. Mean CI during the period of study is 3.24. This is also supported by the mean degree of collaboration at the percentage of 0.83 .The mean CC observed is 0.4635. Regarding the application of Lotka's Law of authorship productivity in the artificial intelligence literature it proved to be fit for the study.

(Ahmad & Batcha, 2020) analyzed the application of Lotka's law to the research publication, in the field of Dyslexia disease. The data related to Dyslexia were extracted from web of science database, which is a scientific, citation and indexing service, maintained by Clarivate Analytics. A total of 5182 research publications were published by the researchers, in the field of Dyslexia. The study found out that, the Lotka's inverse square law is not fit for this data. The study also analyzed the authorship pattern, Collaborative Index (CI), Degree of Collaboration (DC), Co-authorship Index (CAI), Collaborative Co-efficient (CC), Modified Collaborative Co-efficient (MCC), Lotka's Exponent value, Kolmogorov-Smirnov Test (K-S Test), Relative Growth Rate and Doubling Time.

(Jahina, Batcha, & Ahmad, 2020) study deals a scientometric analysis of 8486 bibliometric publications retrieved from the Web of Science database during the period 2008 to 2017. Data is collected and analyzed using Bibexcel software. The study focuses on various aspect of the quantitative research such as growth of papers (year wise), Collaborative Index (CI), Degree of Collaboration (DC), Co-authorship Index (CAI), Collaborative Co-efficient (CC), Modified Collaborative Co-Efficient (MCC), Lotka's Exponent value, Kolmogorov-Smirnov test (K-S Test).

## 3. Objectives

This paper has following objectives:

(a) To analyse the author productivity patterns in the field of South African Publications on CAD.

(b) To examine the validity of Lotka's law using total counting and straight counting of authors.

(c) To apply Kolmolgorov- Smirnov (K-S) goodness of fittest for the conformity of Lotka's law.

## 4. Methodology

The study retrieved and downloaded the publication data of the South Africa and of most productive authors on CAD Research from the Web of Science database during 1990-19. The Search word '"TS=(Artery Disease, Coronary OR Artery Diseases, Coronary OR Coronary Artery Diseases OR Disease, Coronary Artery OR Diseases, Coronary Artery OR Coronary Arteriosclerosis OR Arterioscleroses, Coronary OR Coronary Arterioscleroses OR Atherosclerosis, Coronary OR Atheroscleroses, Coronary OR Coronary Atheroscleroses OR Coronary Atherosclerosis OR Arteriosclerosis, Coronary OR Ischaemic OR Ischemic OR hardening of the Arteries OR Induration of the Arteries OR Arterial Sclerosis ) AND CU=(South Africa)" was used in 'title', abstract and keyword" search tag and restricted it to the period 1990-19 in 'Period range' search tag for searching South African publication data. The searches were performed with all probabilities and bibliographical details of 1284 research paper collectively contributed by medical scientists published in 468 scientific periodical were collected for application of Lotka's Law.

## 5. Data analysis and Discussion

### 5.1. Authorship Pattern

Table 1 illustrates the overall and five year wise distribution of authorship trend. It is evident from the Table 1 that only 7.34 per cent publications were single authored publications while rest of 92.66 had two or more authors. The maximum number of publications were more than ten authored publications (18.97 %) nearly followed by four authored publications (13.74 %), three authored (12.57 %), two authored (12.41 %) and five authored publications (10.62 %). Six to ten authored publications accounted for 24.56 per cent while more than 10 authored publications accounted for 18.93 per cent.

| Author(s) | Total Research Output (5 Yearly) | | | | | | Total Research Output | |
|---|---|---|---|---|---|---|---|---|
| | 1990-1994 | 1995-1999 | 2000-2004 | 2005-2009 | 2010-2014 | 2015-2019 | Total | % |
| Single | 17 | 20 | 17 | 12 | 18 | 10 | 94 | 7.34 |
| Two | 21 | 33 | 13 | 34 | 30 | 28 | 159 | 12.41 |
| Three | 24 | 18 | 22 | 34 | 34 | 29 | 161 | 12.57 |
| Four | 27 | 23 | 22 | 23 | 43 | 38 | 176 | 13.74 |
| Five | 14 | 20 | 10 | 19 | 30 | 43 | 136 | 10.62 |
| Six | 10 | 18 | 15 | 16 | 28 | 35 | 122 | 9.52 |
| Seven | 5 | 13 | 4 | 19 | 14 | 17 | 72 | 5.62 |
| Eight | 2 | 5 | 5 | 11 | 14 | 19 | 56 | 4.36 |
| Nine | 3 | 1 | 4 | 7 | 4 | 14 | 33 | 2.57 |
| Ten | 1 | 1 | 3 | 5 | 10 | 12 | 32 | 2.49 |
| More than 10 | 1 | 8 | 7 | 15 | 62 | 150 | 243 | 18.93 |
| Total | 125 | 160 | 122 | 195 | 287 | 395 | 1284 | 100.00 |
| % | 9.76 | 12.49 | 9.52 | 15.22 | 22.40 | 30.76 | 100.00 | |

Table 1: Authorship Pattern

## 5.2. Lotka's Law

Lotka's Law is one of the most basic laws of bibliometrics, which deals with the frequency of publication of authors in and given field. The generalized form of Lotka's law can be expressed as:

$Y = (C)$

Where Y is the number of authors with X articles, the exponent n and constant C are parameters to be estimated from a given set of author productivity data.

Lotka's law describes the frequency of publication by authors in a given field. It states that the number of authors making n contribution is about $1/n^2$ on those making one and the proportion of all contributions that make a single contributions, is about 60 percent (Potter, 1981). This means that out of all the authors in a given field, 60 percent will have just one publication and 15 percent will have two publications. 7 percent of authors will have three publications and so on. According to Lotka's law of scientific productivity, only six percent the authors in a field will produce more than 10 articles.

While theoretical Lotka's value is á = 2.00

Theoretical value of 'n' = 2.54 is matched with table value of R. Rosseau for getting C.S. value 0.7539.

**Table 2: Lotka's Constant Value and Present Study**

| Constant Value of Present Study | n Value |
|---|---|
| 0.7539 | 2.54 |
| Lotka's Constant Value | n Value |
| 0.6079 | 2 |

D-Max Value Present Study

0.1050

To test the goodness of fittest, weather the observed author productivity distribution is not significantly different from theoretical distribution. K-S test applied. According to this test, the maximum deviation is observed and estimated value D-Max is calculated follows:

$D_{Max} = F(x) - E_n(x)$ á = 2.54

Theoretical Value of $C = 0.7539 \text{ Fe+} = 0.7539 \left( \dfrac{1}{x^{2.54}} \right)$

D-Max = 0.1050

Critical Value at 0.01 level of significance = $\dfrac{2.54}{\sqrt{16006}} = 0.0200$

**Table 3: Lotka's Law on CAD**

| x | y | X=Log x | Y=Log y | XY | X² |
|---|---|---------|---------|-----|-----|
| 1 | 8654 | 0.00000 | 3.93722 | 0.00000 | 0.00000 |
| 2 | 3748 | 0.30103 | 3.57380 | 1.07582 | 0.09062 |
| 3 | 880 | 0.47712 | 2.94448 | 1.40488 | 0.22764 |
| 4 | 566 | 0.60206 | 2.75282 | 1.65736 | 0.36248 |
| 5 | 252 | 0.69897 | 2.40140 | 1.67851 | 0.48856 |
| 6 | 169 | 0.77815 | 2.22789 | 1.73363 | 0.60552 |
| 7 | 1112 | 0.84510 | 3.04610 | 2.57426 | 0.71419 |
| 8 | 191 | 0.90309 | 2.28103 | 2.05998 | 0.81557 |
| 9 | 175 | 0.95424 | 2.24304 | 2.14040 | 0.91058 |
| 10 | 79 | 1.00000 | 1.89763 | 1.89763 | 1.00000 |
| 11 | 44 | 1.04139 | 1.64345 | 1.71148 | 1.08450 |
| 12 | 32 | 1.07918 | 1.50515 | 1.62433 | 1.16463 |
| 13 | 13 | 1.11394 | 1.11394 | 1.24087 | 1.24087 |
| 14 | 16 | 1.14613 | 1.20412 | 1.38008 | 1.31361 |
| 15 | 16 | 1.17609 | 1.20412 | 1.41615 | 1.38319 |
| 16 | 14 | 1.20412 | 1.14613 | 1.38008 | 1.44990 |
| 17 | 10 | 1.23045 | 1.00000 | 1.23045 | 1.51400 |
| 18 | 5 | 1.25527 | 0.69897 | 0.87740 | 1.57571 |
| 19 | 5 | 1.27875 | 0.69897 | 0.89381 | 1.63521 |
| 20 | 3 | 1.30103 | 0.47712 | 0.62075 | 1.69268 |
| 21 | 4 | 1.32222 | 0.60206 | 0.79606 | 1.74826 |
| 22 | 2 | 1.34242 | 0.30103 | 0.40411 | 1.80210 |
| 23 | 2 | 1.36173 | 0.30103 | 0.40992 | 1.85430 |
| 25 | 2 | 1.39794 | 0.30103 | 0.42082 | 1.95424 |
| 26 | 1 | 1.41497 | 0.00000 | 0.00000 | 2.00215 |
| 27 | 2 | 1.43136 | 0.30103 | 0.43088 | 2.04880 |
| 29 | 2 | 1.46240 | 0.30103 | 0.44023 | 2.13861 |
| 31 | 1 | 1.49136 | 0.00000 | 0.00000 | 2.22416 |
| 33 | 1 | 1.51851 | 0.00000 | 0.00000 | 2.30588 |
| 35 | 1 | 1.54407 | 0.00000 | 0.00000 | 2.38415 |
| 50 | 1 | 1.69897 | 0.00000 | 0.00000 | 2.88650 |
| 53 | 1 | 1.72428 | 0.00000 | 0.00000 | 2.97313 |
| 68 | 1 | 1.83251 | 0.00000 | 0.00000 | 3.35809 |
| 114 | 1 | 2.05690 | 0.00000 | 0.00000 | 4.23086 |
| | 16006 | 39.9858 | 40.1046 | 31.4999 | 53.1807 |

The theoretical value of C as 0.7539 for á = 2.54 is taken from the book 'Power Laws in the Information Production Process: Lotkaian Informetrics' by Egghe (2005). The K-S test is applied for the fitness of Lotka's law fits to the South African CAD research output. Result indicates that the value of D- Max, i.e. 0.1050 determined with Lotka's exponent á = 2.54 for CAD which is not close determined with the Lotka's exponent á = 2 than the critical value determined at the 0.01 level of significance, i.e., 0.0200. Thus, distribution frequency of the authorship follows the exact Lotka's inverse law with the exponent á = 2. The modified form of the inverse square law, i.e., inverse power law with á and C parameters as 2.54 and 0.7539 for CAD literature is applicable and appears to provide a good fit.

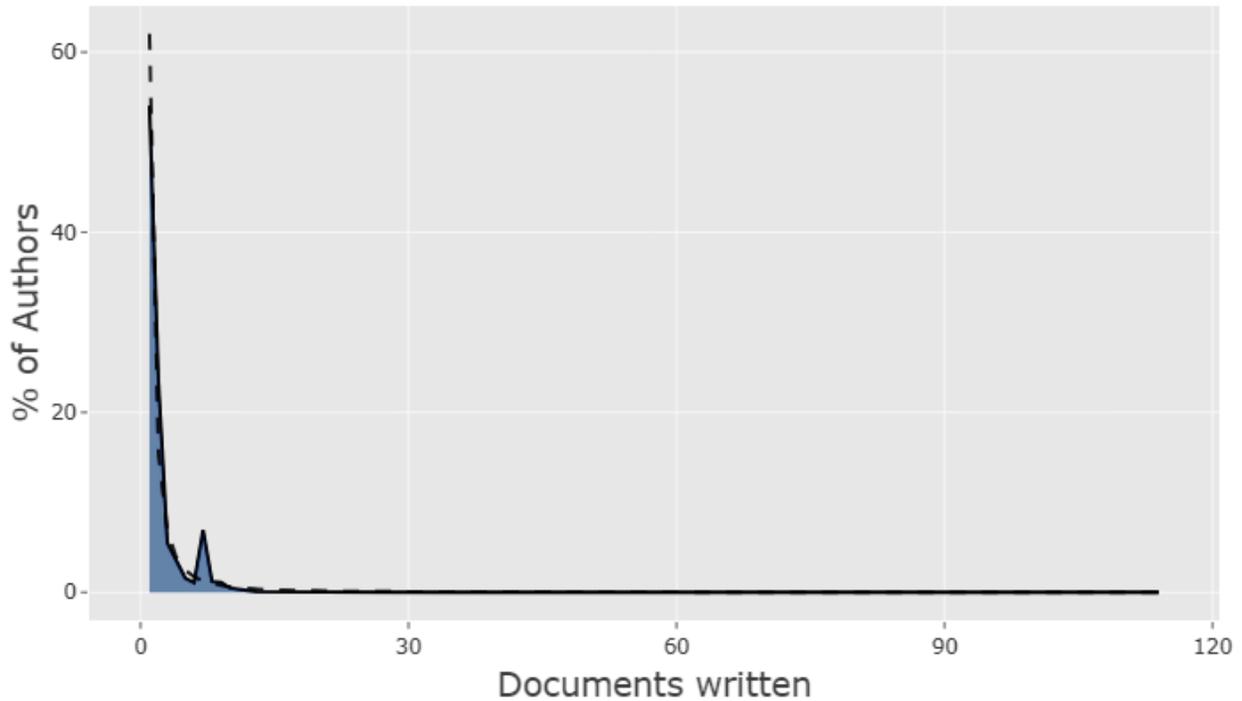

Table 4: K-S Test of Observed and Expected Distribution of Authors

| x | y | Observed=yx/Σxy | Value=Σ(yx/Σyx) | Expected Freq | Value of Freq/Cum | Diff(D) |
|---|---|---|---|---|---|---|
| 1 | 8654 | 0.54067 | 0.540672 | 0.753900 | 0.7539 | -0.2132 |
| 2 | 3748 | 0.23416 | 0.774834 | 0.129182 | 0.8831 | **0.1050** |
| 3 | 880 | 0.05498 | 0.829814 | 0.046031 | 0.9291 | 0.0089 |
| 4 | 566 | 0.03536 | 0.865176 | 0.022135 | 0.9512 | 0.0132 |
| 5 | 252 | 0.01574 | 0.880920 | 0.012545 | 0.9638 | 0.0032 |
| 6 | 169 | 0.01056 | 0.891478 | 0.007888 | 0.9717 | 0.0027 |
| 7 | 1112 | 0.06947 | 0.960952 | 0.005328 | 0.9770 | 0.0641 |
| 8 | 191 | 0.01193 | 0.972885 | 0.003793 | 0.9808 | 0.0081 |
| 9 | 175 | 0.01093 | 0.983819 | 0.002811 | 0.9836 | 0.0081 |
| 10 | 79 | 0.00494 | 0.988754 | 0.002150 | 0.9858 | 0.0028 |
| 11 | 44 | 0.00275 | 0.991503 | 0.001687 | 0.9874 | 0.0011 |
| 12 | 32 | 0.00200 | 0.993502 | 0.001352 | 0.9888 | 0.0006 |
| 13 | 13 | 0.00081 | 0.994315 | 0.001102 | 0.9899 | -0.0003 |
| 14 | 16 | 0.00100 | 0.995314 | 0.000913 | 0.9908 | 0.0001 |
| 15 | 16 | 0.00100 | 0.996314 | 0.000766 | 0.9916 | 0.0002 |
| 16 | 14 | 0.00087 | 0.997189 | 0.000650 | 0.9922 | 0.0002 |
| 17 | 10 | 0.00062 | 0.997813 | 0.000557 | 0.9928 | 0.0001 |
| 18 | 5 | 0.00031 | 0.998126 | 0.000482 | 0.9933 | -0.0002 |
| 19 | 5 | 0.00031 | 0.998438 | 0.000420 | 0.9937 | -0.0001 |
| 20 | 3 | 0.00019 | 0.998625 | 0.000368 | 0.9941 | -0.0002 |
| 21 | 4 | 0.00025 | 0.998875 | 0.000325 | 0.9944 | -0.0001 |
| 22 | 2 | 0.00012 | 0.999000 | 0.000289 | 0.9947 | -0.0002 |
| 23 | 2 | 0.00012 | 0.999125 | 0.000258 | 0.9949 | -0.0001 |
| 25 | 2 | 0.00012 | 0.999250 | 0.000209 | 0.9951 | -0.0001 |
| 26 | 1 | 0.00006 | 0.999313 | 0.000189 | 0.9953 | -0.0001 |
| 27 | 2 | 0.00012 | 0.999438 | 0.000172 | 0.9955 | 0.0000 |
| 29 | 2 | 0.00012 | 0.999563 | 0.000143 | 0.9956 | 0.0000 |
| 31 | 1 | 0.00006 | 0.999625 | 0.000121 | 0.9958 | -0.0001 |
| 33 | 1 | 0.00006 | 0.999688 | 0.000103 | 0.9959 | 0.0000 |
| 35 | 1 | 0.00006 | 0.999750 | 0.000089 | 0.9960 | 0.0000 |
| 50 | 1 | 0.00006 | 0.999813 | 0.000036 | 0.9960 | 0.0000 |
| 53 | 1 | 0.00006 | 0.999875 | 0.000031 | 0.9960 | 0.0000 |
| 68 | 1 | 0.00006 | 0.999937 | 0.000016 | 0.9960 | 0.0000 |
| 114 | 1 | 0.00006 | 1.000000 | 0.000004 | 0.9960 | 0.0001 |
| Total | 16006 | | Present Study D.Max = 0.1050 | | | |

### 6. Findings and Conclusion

Generally Lotka's Law describes the frequency of publications by authors in a given subject or discipline. In this paper, an attempt has been made to study the applicability of the Lotka's Law to the South African publications on a subject or a discipline of Coronary Artery Disease. A K-S test is applied for the fitness of Lotka's law fits to the CAD data. Result indicates that the value of D-Max, i.e., 0.1050 determined with Lotka's exponent, á = 2.54 and is not close to the D-Max value with the Lotka's exponent á = 2 than the critical value determined at the 0.01 level of significance, i.e., 0.0200. Thus, distribution frequency of the authorship follows the exact Lotka's Inverse Law with the exponent á = 2. However, the modified form of the inverse square law, i.e. Inverse Power Law with á and C parameters as 2.54 and 0.7539 is not applicable and does not appear to provide a good fit. Lotka's law of author productivity is regarded as one of the classical laws of bibliometric. The present study showed that Lotka' generalized law is not applicable to CAD literature. K-S test is applied to verify the applicability of Lotka's law of scientific productivity. The statistical tests show that the Lotka's law in its generalized form does not fit the author productivity distribution pattern prepared for the straight count and for the contribution of complete count of the Coronary Artery Disease research Output.